# Modelling Weak-Coherent CV-QKD Systems Using a Classical Simulation Framework

**Sören Kreinberg[1], Igor Koltchanov[1], Piotr Novik[2], Saleem Alreesh[1], Fabian Laudenbach[3], Christoph Pacher[3], Hannes Hübel[3], André Richter[1,*]**

[1] *VPIphotonics GmbH, Carnotstr. 6, 10587 Berlin, Germany*

[2] *VPI Development Center, ul. Filimonova 15-50831, 220037 Minsk, Belarus*

[3] *Austrian Institute of Technology GmbH, Donau-City-Str. 1, 1220 Vienna, Austria*

* *Tel: +49 30 398058-0, Fax: +49 30 398958-58, e-mail: andre.richter@vpiphotonics.com*

**ABSTRACT**

Due to their compatibility to existing telecom technology, continuous variable (CV) weak coherent state protocols are promising candidates for a broad deployment of quantum key distribution (QKD) technology. We demonstrate how an existing simulation framework for modelling classical optical systems can be utilized for simulations of weak-coherent CV-QKD links. The quantum uncertainties for the measured characteristics of coherent signals are modelled in the electrical domain by shot noise, while a coherent signal in the optical domain is described by its quadrature components. We simulate various degradation effects such as attenuation, laser RIN, Raman noise (from classical channels in the same fibre), and device imperfections and compare the outcome with analytical theory. Having complemented the physical simulation layer by the post-processing layer (reconciliation and privacy amplification), we are able to estimate secure key rates from simulations, greatly boosting the development speed of practical CV-QKD schemes and implementations.

## 1. INTRODUCTION

For secure communication, secure key exchange is required. Currently, the key exchange is often performed via asymmetric cryptography. Asymmetric cryptography relies on the assumption that certain trapdoor functions are hard to invert. Future advances in computing hardware or algorithms may invalidate this assumption. In this sense, advances in quantum computing pose a dangerous threat. A promising solution to this dilemma is offered by quantum key distribution (QKD), which makes use of the quantum no-cloning theorem to ensure the secrecy of key exchange on the basis of quantum physical and information theoretical principles. QKD allows to reliably estimate the maximum amount of information an eavesdropper might have gained. Either this information can be reduced to zero during post-processing (without discarding the entire mutual information), or the protocol aborts. The huge variety of QKD protocols can be divided into discrete variable (DV) and continuous variable (CV) protocols depending on whether the eigenvalue spectrum of the quantum mechanical measurement operator is discrete or continuous. On the field of CV-QKD [1], prepare-and-measure protocols using weak coherent states are very promising candidates for a broad deployment, as they use the electric field quadratures for encoding the quantum information (quadrature amplitude modulation, QAM). QAM is everyday business in optical telecommunication and experimental implementations of CV-QKD using standard telecom hardware have been demonstrated [2], [3].

This work focusses on demonstrating how VPItransmissionMaker Optical Systems, a framework for simulating optical communication links on the system level, can be used to model prepare-and-measure weak-coherent CV-QKD in order to analyse and optimise parameters of the transmission link or the QKD protocol.

## 2. SIMULATION FRAMEWORK

Coherent states are often referred to as the "most classical" quantum states and can be understood as electric fields or laser pulses. Consequently, the quadratures $q$ and $p$ of the corresponding quantum field can be described by the real and imaginary part of a complex-valued electric field. VPItransmissionMaker Optical Systems treats time-varying electric fields as a time-discrete sequence of phasors relative to a reference frequency. Shot noise occurring during detection of a coherent state using a photodiode can be simulated in this framework by drawing a sample from the corresponding Poissonian distribution. Without further adjustments it is therefore possible to use VPItransmissionMaker Optical Systems for simulating prepare-and-measure weak-coherent CV-QKD systems.

## 3. COMPARING SIMULATION RESULTS TO ANALYTICAL THEORY

Recently, Laudenbach et al. [4] published a comprehensive theoretical analysis of different noise sources degrading Gaussian-modulated CV-QKD. For demonstrating the possibilities of our simulation framework, we constructed a basic scenario for which we simulated the effects of the noise sources on Gaussian-modulated CV-QKD as discussed in [4]. The corresponding schematic (Fig. 1) consists of Alice's coherent state preparation, transmission via a quantum channel, Bob's receiver and post-processing for estimating important link parameters.

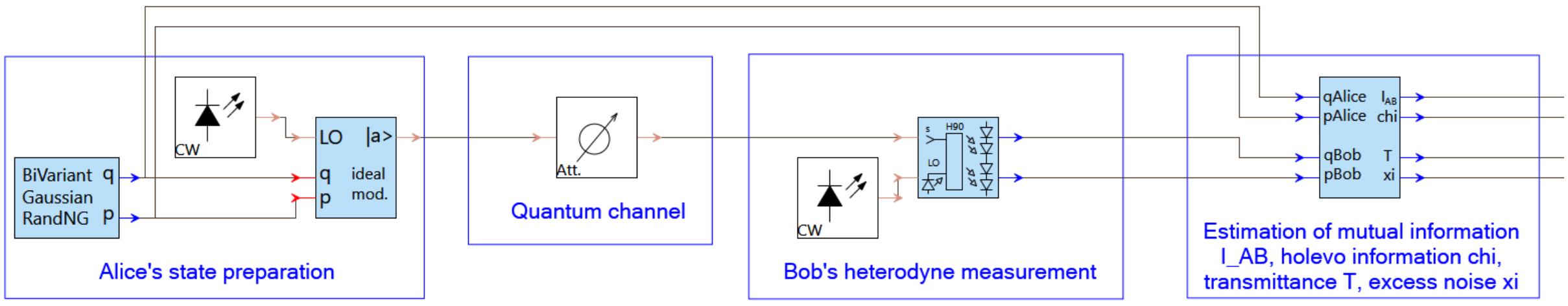

*Figure 1. Simulation schematic for a basic Gaussian CV-QKD demo.*

These are the effective transmittance $T$, excess noise $\xi$, mutual information between Alice and Bob $I_{AB}$ and the Holevo upper bound of Eve's knowledge over Bob's measurement result $\chi_{EB}$. For reducing complexity we simulate exactly one signal sample per quantum symbol and set the linewidths and relative detuning of the carrier laser and local oscillator (LO) laser to zero. The effects of the most important noise sources on the excess noise parameter $\xi$ are shown in Fig. 2 and Fig. 3. The simulation results (blue) perfectly match the theoretical predictions (orange) within the margin of statistical uncertainty.

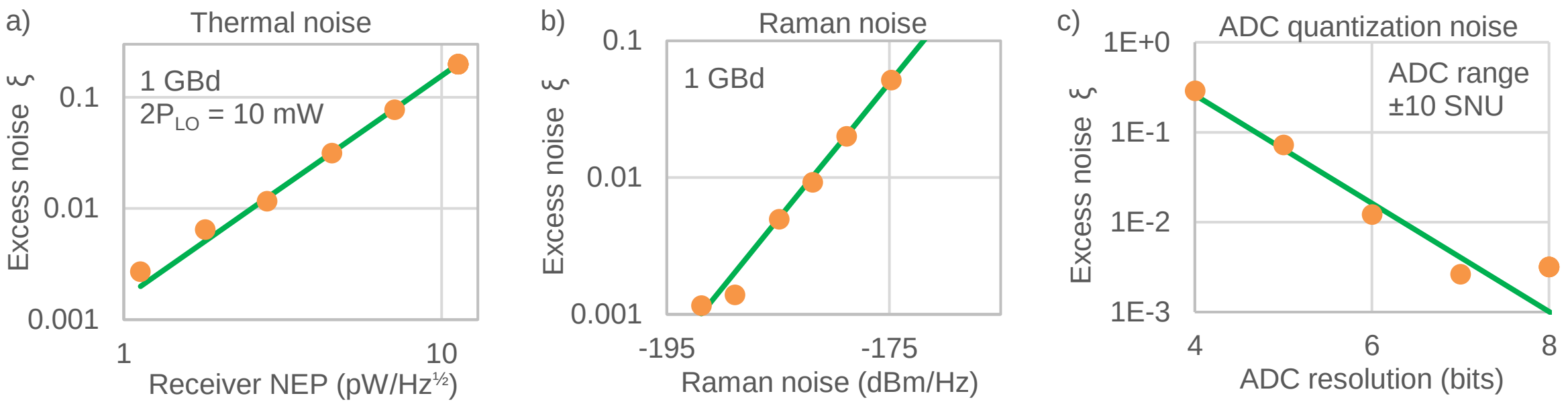

*Figure 2. Comparison of calculated (green curve) and measured (orange dots) excess noise $\xi$ for different noise sources. Panel a) depicts $\xi$ in dependence of the receiver noise-equivalent power (NEP). Panel b) depicts $\xi$ in dependence of the Raman noise spectral density caused by a co-existing classical channel in the same fibre. Panel c) shows the excess noise caused by ADC quantization in dependence of the ADC resolution.*

Additionally, we investigated the effect of important parameters such as receiver thermal noise in terms of optical noise-equivalent power (NEP), signal attenuation in terms of channel transmittance $T_{ch}$ and modulation variance $V_{mod}$ on the mutual information $I_{AB}$ and the Holevo bound $\chi_{EB}$. The parameter variations were performed around an operation point with $V_{mod}$ = 5, quantum efficiency $\eta$ = 0.7, NEP = 2.82 pW/Hz½, $T_{ch}$ = 0.3. The symbol rate was 1 GBd and Bob's LO laser had an optical power of $2P_{LO}$ = 10 mW. The results are shown in Fig. 4. Because increasing the receiver NEP in Fig. 4a also increases the excess noise $\xi$, the mutual information between Alice and Bob decreases slightly and the Holevo bound of Eve's information on Bob's measurement result increases strongly. Fig. 4b illustrates how both mutual information $I_{AB}$ and Holevo bound $\chi_{EB}$ increase with increased transmittance. For low transmittance, the Holevo bound can be higher than the mutual information, because the losses are attributed to Eve manipulating the channel. For very low modulation variance $V_{mod}$, the Holevo bound $\chi_{EB}$ is larger than the mutual information $I_{AB}$. For practical applications it is important to find the modulation variance that delivers a good relative information margin in favour of Alice and Bob, which for the given case in Fig. 4c is approximately around 5.

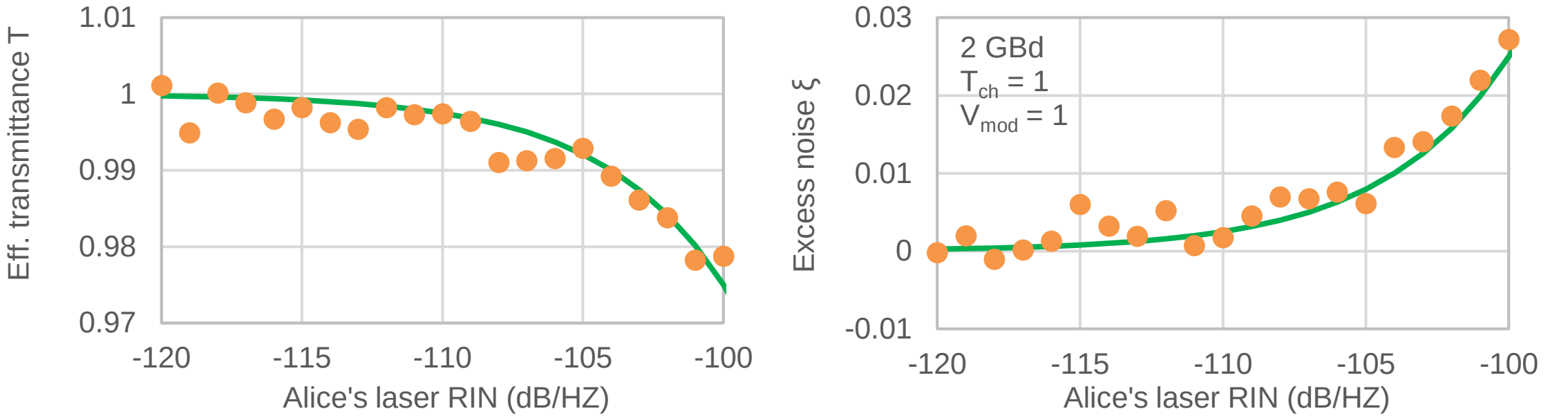

*Figure 3. Comparison of calculated (green curve) and measured (orange dots) effective transmittance T and excess noise $\xi$ for different densities of Alice's laser relative intensity noise (RIN).*

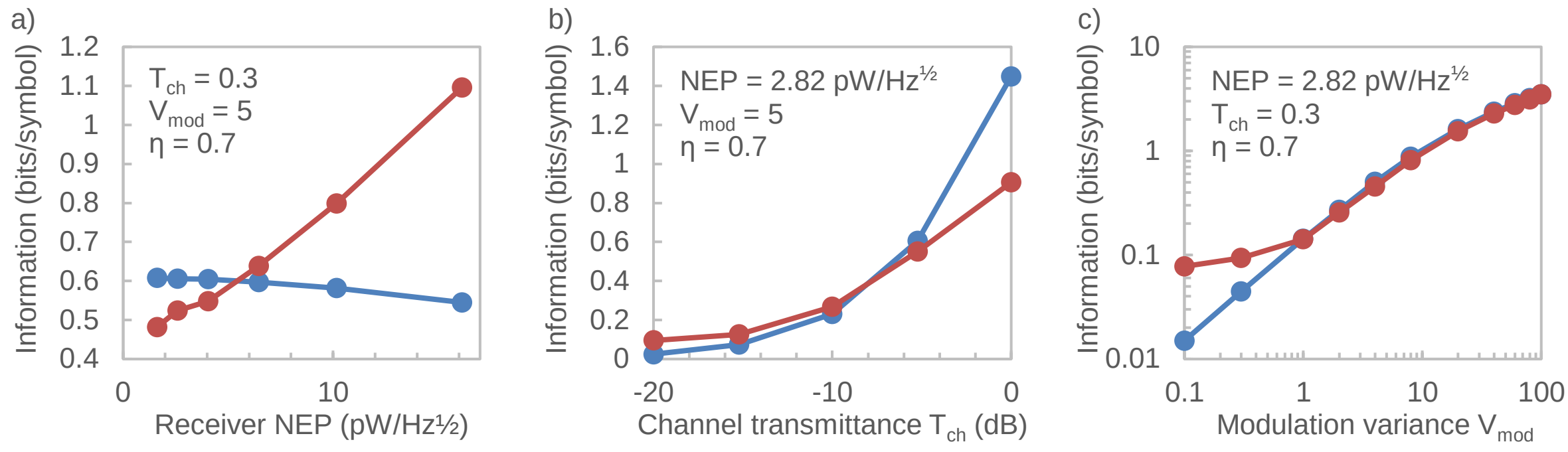

*Figure 4. Mutual information $I_{AB}$ between Alice and Bob (blue) and Holevo upper bound $\chi_{EB}$ of Eve's information on Bob's state (red). Photodiode thermal noise, channel transmittance and modulation variance are varied around $V_{mod}$ = 5, $\eta$ = 0.7, NEP = 2.82 pW/Hz½, $T_{ch}$ = 0.3.*

## 4. MATCHED FILTERING FOR OPTIMIZED SECRET KEY RATE

Simulation of realistic CV-QKD experiments requires modelling of arbitrary symbol pulse shapes and thus multiple samples per symbol. On the receiver side non-trivially shaped pulses need to be analysed to gain as much information on the quadratures as possible from the incoming symbol pulses. Moreover, the quadratures need to be converted to shot noise units (SNU). The conversion to SNU can be achieved by running a simulation script executing the calibration sequence described in [4], whereas the analysis of incoming pulses is performed by matched filtering. In the realized schematic each symbol was described by 16 samples at a symbol rate of 250 MBd. The pulses had $sin^4$- shape and were filtered on the receiver side using a Gaussian low pass filter. The Gaussian filter matches the pulse shape quite well. A series of simulations was run to investigate how the effective transmittance $T$ and excess noise $\xi$ depend on the low pass filter bandwidth. The results in Fig 5 show that the effective transmittance $T$ has a pronounced optimum when the filter bandwidth is approximately 90 % of the symbol rate, whereas the excess noise $\xi$ turns out to be quite robust against a bandwidth mismatch and only increases for very small bandwidths. For isolated analysis of the noise added by filter bandwidth, all other possible excess noise sources were disabled in simulations.

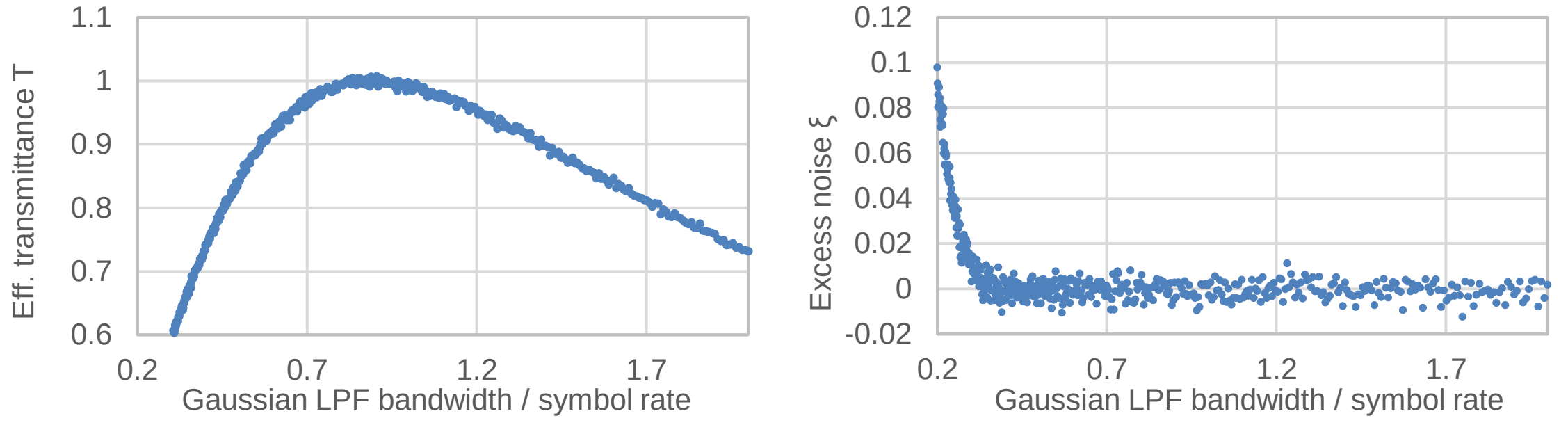

*Figure 5. Effective transmittance $T$ and excess noise parameter $\xi$ in dependence on the ratio of the low pass filter bandwidth and symbol rate.*

## 5. PHASE RECOVERY IS ALMOST EXCESS NOISE-FREE

Early CV-QKD proposals [2], [3] suggested that Bob uses light from the same local oscillator laser as Alice does. It was discovered later, that this approach opens a security breach [5] and that a true local oscillator (or "local local oscillator") on Bob's side is required. The phase diffusion between Alice's and Bob's laser (due to non-zero linewidth) as well as phase variation due to changing optical path length can be compensated if Alice transmits a pilot tone as a phase reference over the same fibre. This pilot tone delivers the phase information to Bob who is thus able to recover the original phase of the quantum signal via digital signal processing. Schemes involving time- and polarization multiplexing have been proposed to transmit the pilot tone through the fibre alongside the quantum signal [6], [7]. A very recent proposal suggests offsetting the pilot tone by 1 GHz and sending it orthogonally polarized to the quantum signal [8]. It was reported, that the phase recovery technique adds almost no excess noise to the quantum channel. We implemented a very similar scheme to the one suggested in [6] and were indeed able to confirm this finding. The simulated excess noise in dependence of modulation variance and signal attenuation is depicted in Fig. 6. Again, for investigating purely the excess noise added by the phase recovery, we disabled all other possible excess noise sources during simulations.

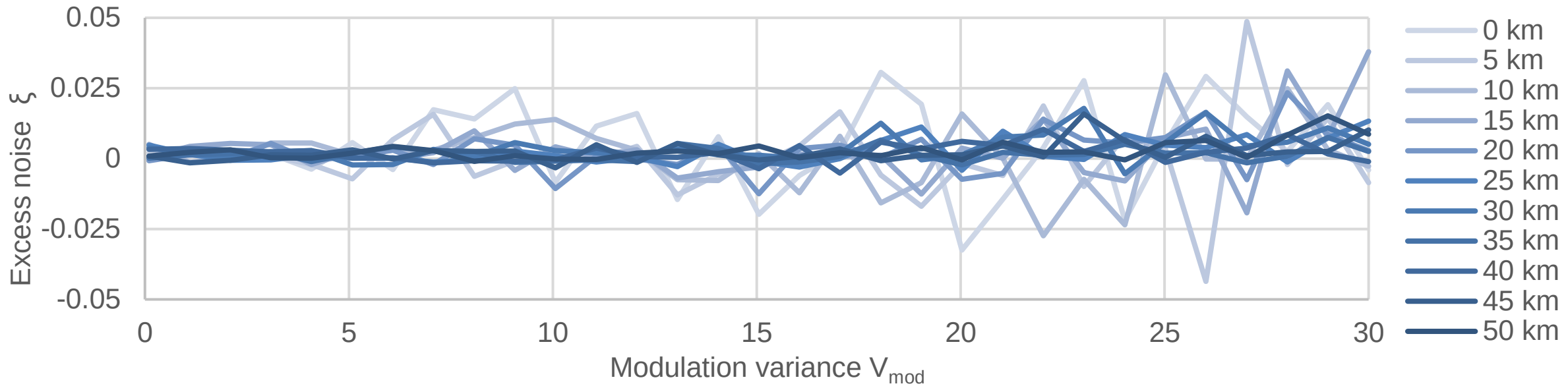

*Figure 6. Excess noise $\xi$ in dependence of the modulation variance $V_{mod}$ and the signal attenuation. For the simulations, a reconciliation efficiency of 95% was assumed. The receiver NEP was set to zero, the fibre attenuation was 0.2 dB/km. For every data point, $2^{20}$ symbols were used for parameter estimation.*

## 6. SECRET KEY EXTRACTION

We implemented the reconciliation and privacy amplification scheme proposed in Fig. 3 of ref. [9] as independent toolchain. Using an R = 0.1 MET-LDPC code [10] of length 200 000, 8D rotation and heterodyne detection, we obtained a secret key fraction of 0.1 bits/symbol for $\eta$ = 0.7, $T_{ch}$ = 0.35, $V_{mod}$ = 1.35 and $\xi$ = 0.0116, if 10 % of the transmitted symbols are used for parameter estimation. Current efforts focus on preparing a dense palette of codes and on integrating the toolchain into our simulation framework.

## 7. CONCLUSIONS AND OUTLOOK

Using VPItransmissionMaker Optical Systems, we demonstrated that a classical optical simulation framework can be used for modelling weak-coherent prepare-and-measure CV-QKD systems. The simulations of various noise sources fit the analytical predictions. As was shown, it is possible to test the suitability of receivers, filtering schemes and digital signal processing solutions (e.g. for phase recovery). The flexibility to selectively turn on and off physical effects and noise sources opens up research possibilities not accessible in experiment. Current work focusses on integrating our reconciliation and privacy amplification toolchain into the simulation framework, with the ultimate goal of providing a comprehensive tool for precise secret key rate prediction.

## ACKNOWLEDGEMENTS

This work is part of the UNIQORN project and has received funding from the European Union's Horizon 2020 research and innovation programme under grant agreement No 820474.